\documentclass[aps,twocolumn,floatfix,superscriptaddress,pra]{revtex4-1}
\usepackage{amssymb}
\usepackage{graphicx}
\usepackage{dcolumn}
\usepackage{amsmath}
\usepackage{bm}
\usepackage{yhmath}
\usepackage{textcomp}
\usepackage{epstopdf}
\usepackage{color}
\usepackage{ulem}
\usepackage[toc,page,title,titletoc,header]{appendix}
\usepackage[colorlinks,
            linkcolor=blue,
            anchorcolor=blue,
            citecolor=blue
            ]{hyperref}
\usepackage{natbib}

\begin{document}

\preprint{APS/123-QED}

\title{Superfluid properties of a honeycomb dipolar supersolid}

\author{Albert Gallem\'i}
\affiliation{Institut f\"ur Theoretische Physik, Leibniz Universit\"at Hannover, 30167 Hannover, Germany}
\author{Luis Santos}%
\affiliation{Institut f\"ur Theoretische Physik, Leibniz Universit\"at Hannover, 30167 Hannover, Germany}%

\date{\today}


\begin{abstract}

Recent breakthrough experiments on dipolar condensates have reported the 
creation of supersolids, including two-dimensional arrays of quantum droplets. 
Droplet arrays are, however, not the only possible non-trivial density 
arrangement resulting from the interplay of mean-field instability and 
quantum stabilization. Several other possible density patterns may occur in 
trapped condensates at higher densities, including the so-called honeycomb 
supersolid, a phase that exists, as it is also the case of a triangular 
droplet supersolid, in the thermodynamic limit. We show that compared to 
droplet supersolids, honeycomb supersolids have a much-enhanced superfluid 
fraction while keeping a large density contrast, and constitute in this 
sense a much better dipolar supersolid. However, in contrast to droplet 
supersolids, quantized vortices cannot be created in a honeycomb supersolid 
without driving a transition into a so-called labyrinthic phase. We show 
that the reduced moment of inertia, and with it the superfluid fraction, 
can be however reliably probed by studying the dynamics following a 
scissors-like perturbation.

\end{abstract}

\maketitle


\section{Introduction}
\label{sec:introduction}


The recent realization of the supersolid phase in ultracold dipolar quantum 
gases~\cite{Tanzi2019,Bottcher2019,Chomaz2019} opens intriguing questions about the 
superfluidity in these systems~\cite{Leggett1970}. The interplay between 
mean-field interactions, quantum fluctuations, and external confinement results 
in a rich landscape of possible density patterns, depending on the 
atom-atom interactions, the density, and 
the external trap~\cite{Hertkorn2021,Zhang2019,Zhang2021,Poli2021}. Of these possible patterns, 
droplet arrays have been the most intensively explored phase. Dipolar droplet supersolids~(D-SS) 
have been realized in recent years both in one-~\cite{Bottcher2019,Chomaz2019,Tanzi2019} 
and two-dimensional~\cite{Norcia2021,Bland2022} arrangements. In two dimensions, quantum droplets arrange in a triangular crystalline-like order, surrounded by a low-density region that ensures phase coherence amongst the different droplets. 


Whereas most of the predicted density patterns demand an external confinement, in addition to the D-SS, only another modulated phase may occur in homogeneous space~\cite{Zhang2019}, namely the so-called honeycomb supersolid~(H-SS). This pattern consists of a condensate containing hexagonally-arranged density holes, constituting the complementary pattern to the triangular droplet array. The observation of the H-SS demands significantly larger densities than the D-SS. Although this is still an important limitation in current experiments based on magnetic atoms, the problem could be circumvented 
with a tight confinement~\cite{Zhang2021,Poli2021}, and could be much less relevant in future experiments with condensates of polar molecules~\cite{Schmidt2022}.


Although recent experiments have revealed the expected presence of 
two different Goldstone modes associated with the phase coherence and the crystalline order~\cite{Tanzi2019b,Guo2019,Natale2019}, a direct clear proof of superfluid effects in dipolar supersolids is still lacking. 
Whereas in standard condensates the study of the scissors mode frequency
provides a clear information about the reduction of the moment of inertia, and with it about superfluidity~\cite{GueryOdelin1999,Marago2000}, recent studies have revealed that a more subtle analysis 
of the scissors response is necessary in D-SSs~\cite{Tanzi2021,Norcia2022,Roccuzzo2022}.
An alternative would be provided by the observation of quantized vortices. Recent studies 
have shown that vortices may be robustly created in D-SSs, where they occupy the inter-droplet interstitials~\cite{Roccuzzo2020,Gallemi2020,Sindik2022}. Vortices 
have not yet been observed in a D-SS, but the recent experimental creation of vortices in an unmodulated dipolar condensate~\cite{Klaus2022} opens optimistic 
perspectives in this direction.


These recent works on dipolar density patterns and on superfluid effects in D-SSs 
lead to intriguing questions about the superfluidity of other density patterns, and 
very especially of the H-SS, and about how the 
superfluidity of this phase may be experimentally revealed. This paper is devoted to these 
questions. We show that the H-SS has for a large contrast of the density modulation a much larger superfluid fraction than the D-SS. However, 
contrary to the D-SS, the H-SS is not robust when vortices nucleate, rather transitioning into a labyrinthic phase. Fortunately, the large superfluid fraction of the H-SS could 
be revealed under realistic conditions by a careful analysis of the scissors response.


The paper is organized as follows. In Sec.~\ref{sec:groundstate}, we 
introduce the formalism employed, and review the possible ground-state density patterns.
Section~\ref{sec:rotation} is devoted to the study of the density patterns under rotation. 
In Sec.~\ref{sec:momentinertiasec}, we analyze the moment of inertia of the different patterns, paying special attention to the H-SS. Section~\ref{sec:scissors} discusses in detail how 
the response to a scissors perturbation may be employed to provide a good estimation 
of the reduction of the moment of inertia in the H-SS, revealing its large superfluid fraction. 
Our conclusions are summarized in Sec.~\ref{sec:conclusions}.



\begin{figure}[t!]
\centering
\includegraphics[width=\linewidth]{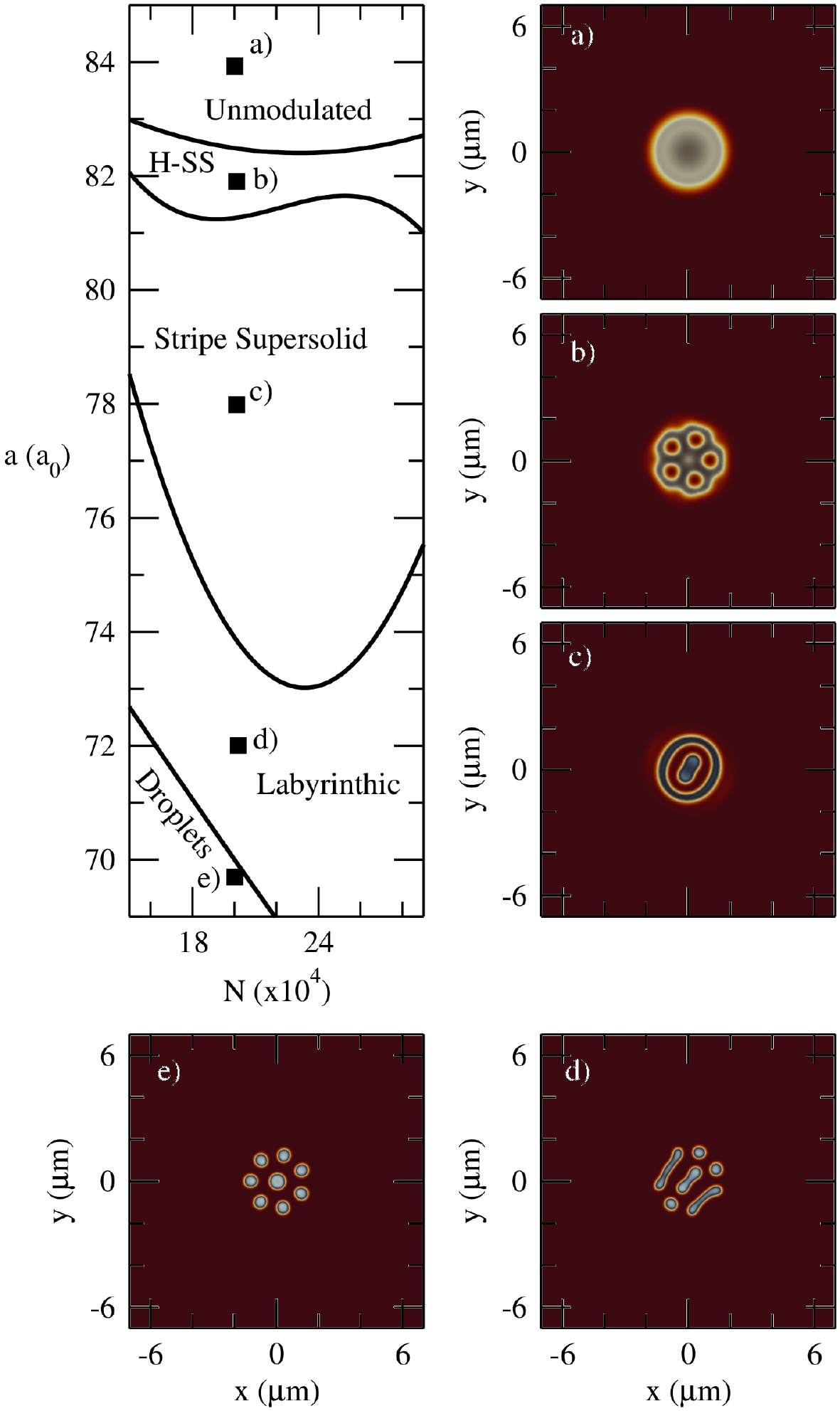}
\caption{(Top left) Ground-state phase diagram as a function of the scattering length and 
the atom number, for $^{162}$ Dy atoms confined in a harmonic 
trap with frequencies $\omega_{\rm x,y,z}=2\pi\times (200,200,400)$. The  
panels (a)-(e) show the density pattern in the $xy$ plane for $N=2\times10^5$ and 
$a/a_0=70,72,78,82$ and $84$, which correspond to the droplet, labyrinthic, stripe,
H-SS, and unmodulated patterns, respectively.}
\label{fig:phase_diagram}
\end{figure}



\section{Ground-state patterns}
\label{sec:groundstate}

In the following, we consider a condensate of $N$ bosons with mass $m$ and 
magnetic dipole moment $\mu$ oriented along the $z$ direction. A similar 
physics is expected for the case of electric dipoles, as it is in particular 
the case of polar molecules~\cite{Schmidt2022}. The physics of a dipolar condensate 
is given by the interplay between contact interactions, characterized by the 
$s$-wave scattering length $a$, dipolar interactions, and (despite of its 
weakly-interacting character) quantum fluctuations. The latter provides the 
stabilization mechanism against mean-field collapse~\cite{Petrov2015}. The physics of a 
quantum-stabilized dipolar condensate is well described by the so-called 
extended Gross-Pitaevskii equation~\cite{Waechtler2016}, 
\begin{eqnarray}
i\hbar \dot\Psi({\bf r}, t)&=& \left [-\frac{\hbar^2}{2m}\nabla^2+V_{\rm trap}({\bf r})+g|\Psi({\bf r}, t)|^2 \right\delimiter 0 \nonumber\\
&+&\int V_{\rm dd}({\bf r}-{\bf r}')|\Psi({\bf r}', t)|^2d{\bf r}'\nonumber\\
&+& \left \delimiter 0\gamma |\Psi({\bf r}, t)|^3 \right ]\Psi({\bf r}, t)\,.
\label{eq:eGPE}
\end{eqnarray}
where $\Psi({\bf r}, t)$ is the condensate wave function.
In Eq.~\eqref{eq:eGPE} $V_{\rm trap}({\bf r})$ is the external 
trapping of the system, $g=4\pi\hbar^2a/m$ is the coupling constant,  
$V_{\rm dd}=\frac{\mu_0\mu^2}{4\pi}\frac{1-3\cos^2\theta}{|{\bf r}-{\bf r}'|^3}$ 
is the dipole-dipole interaction potential, with $\mu_0$ the vacuum magnetic permeability, and $\theta$ the angle between ${\bf r}-{\bf r}'$ and the 
polarization direction. The last term of the equation accounts for the effects of quantum fluctuations, 
the so-called Lee-Huang-Yang correction \cite{Lima2012}, with 
\begin{equation}
\!\!\!\gamma\!=\!\frac{16}{3\sqrt{\pi}}ga^{3/2}
\mbox{Re}\left [\int_0^\pi\!\!\! d\phi\sin\phi[1\!+\!\varepsilon_{\rm dd}(3\cos^2\phi-1)]^{\frac{5}{2}}\! \right ]\! ,
\end{equation}
where $\varepsilon_{\rm dd}=g_{\rm dd}/g$ is the ratio between the dipolar strength $g_{\rm dd}=\mu_0\mu^2/3$ and the 
contact strength $g$. The ground state is obtained by setting in Eq.~\eqref{eq:eGPE} $\Psi({\bf r}, t)=e^{-i\mu t/\hbar} \Psi({\bf r})$, with $\mu$ the chemical potential. Normalizing 
$\int d^3 r |\Psi({\bf r})|^2 = N$, we obtain $\Psi({\bf r})$ using standard imaginary-time evolution techniques.

The interplay between dipole-induced mean-field instability, quantum 
stabilization, and external confinement results in a variety of possible 
ground state phases. If the condensate is mean-field stable, the density 
profile remains unmodulated, presenting (for sufficiently large interactions) 
a standard Thomas-Fermi parabolic form. We denote this phase in the following 
as the unmodulated phase. For a sufficiently low $a$, and small-enough $N$, 
the condensate is mean-field unstable, breaking into quantum droplets, which 
may remain coherently linked (supersolid regime) or eventually disconnected 
(independent-droplet regime). These phases have been experimentally observed 
both in quasi-one- (\cite{Tanzi2019,Bottcher2019,Chomaz2019}) and quasi-2D 
geometries (\cite{Norcia2021,Bland2022}). In the latter case, in the 
thermodynamic limit, the droplet pattern is expected to acquire a triangular 
geometry \cite{Zhang2019,Zhang2021,Bland2022}. Recent works~\cite{Hertkorn2021,Zhang2021} 
have shown, however, that the ground state phase diagram may be significantly 
richer for larger atoms numbers (for fixed external confinement). Particularly 
interesting is the so-called honeycomb phase, which as for the case of the 
triangular droplet array, extends all the way to the thermodynamic limit 
\cite{Zhang2019,Zhang2021}. This phase is characterized by the formation of 
empty regions, which are arranged in an hexagonal pattern. In the presence 
of harmonic confinement, a rich variety of additional patterns may occur, 
including the so-called labyrinthic, stripe, and pumpkin phases \cite{Hertkorn2021}. 

Figure~\ref{fig:phase_diagram}~(top left panel) illustrates these possible 
ground-state phases for the particular case of a condensate of $^{162}$Dy 
atoms (which possess a dipole moment $\mu=10\mu_B$, $\mu_B$ being the Bohr 
magneton). We consider throughout this paper an external harmonic confinement 
\begin{equation}
V_{\rm trap}=\frac{1}{2}m(\omega_{\rm x}x^2+\omega_{\rm y}y^2+\omega_{\rm z}z^2)\,,
\end{equation}
with frequencies $\omega_{\rm x,y}=\omega=2\pi\times 200\,\mathrm{Hz}$ and 
$\omega_{\rm z}=2\pi\times 400\,\mathrm{Hz}$. 
For $N$ up to $3\times10^5$ atoms and $70<a/a_0<84$, with $a_0$ the Bohr 
radius, the system presents different density patterns: unmodulated  
(panel (a)), H-SS ~(panel (b)), stripe phase~(panel (c))~(formed by a pattern of concentric rings), labyrinthic~(panel (d)) 
and droplet array~(panel (e)). The so-called pumpkin pattern only appears in this case
for much larger atom numbers~\cite{Hertkorn2021}.



\begin{figure}[t!]
\centering
\includegraphics[width=\linewidth]{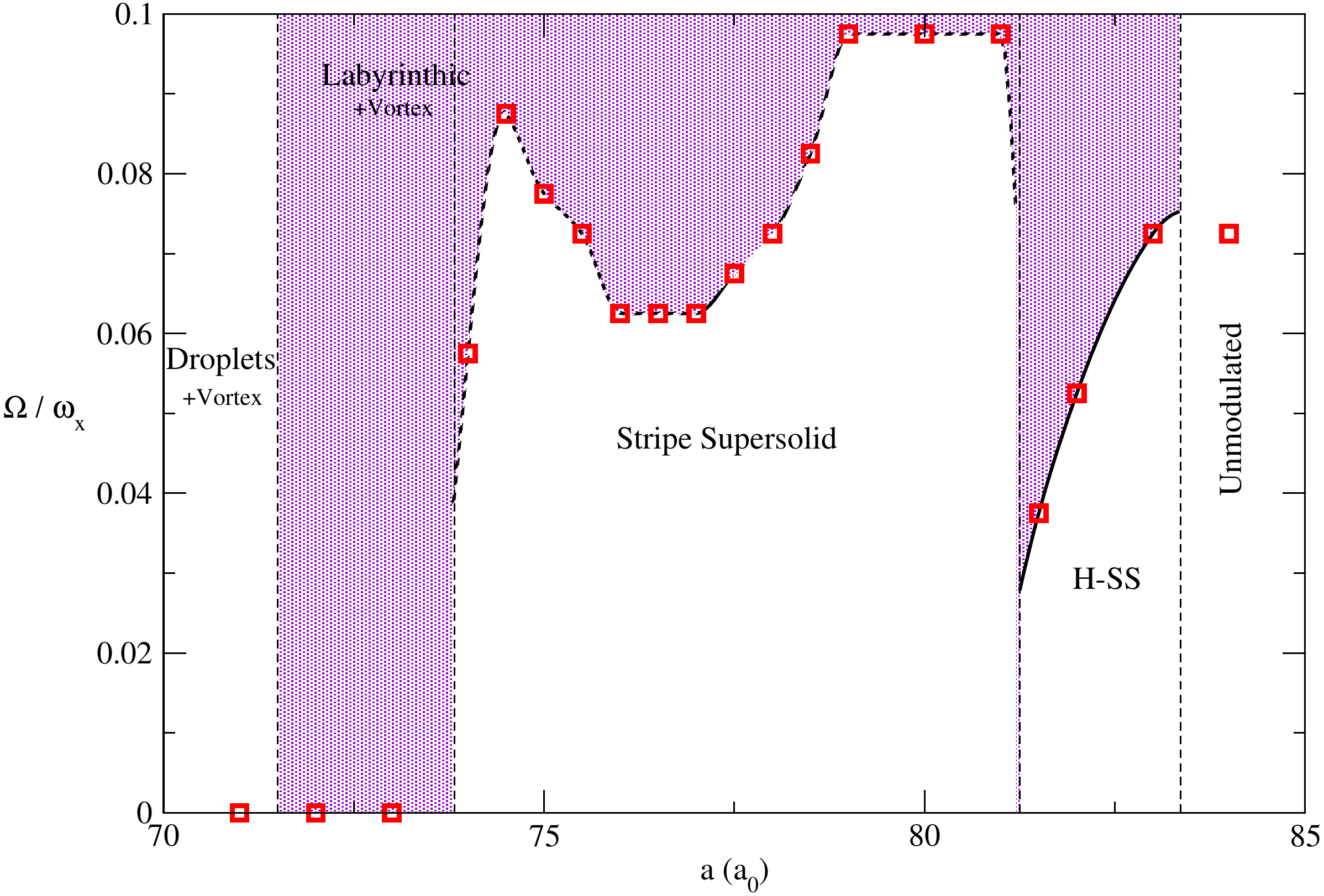}
\caption{Ground-state phase diagram of the system under rotation for 
$N=2\times 10^5$ as a function of the rotation frequency $\Omega$, 
for the same parameters as in Fig.~\ref{fig:phase_diagram}. 
Above a certain critical value of $\Omega$~(represented by the line in the plot 
and the red squares), a vortex nucleates. Note that for the stripe and the H-SS phases 
this results in the breaking of the pattern into a labyrinthic arrangement (shaded region).}
\label{fig:rotation_diagram}
\end{figure}



\section{Robustness of the density pattern under rotation}
\label{sec:rotation}

The creation of quantized vortices constitute a possible way to directly 
probe superfluidity. We consider at this point that the condensate 
is rotated around $z$ with a rotation frequency $\Omega$. We obtain the 
ground state by evaluating the eGPE in the rotating frame, minimizing 
$H-\Omega \hat L_z$, with $\hat L_z$ the angular momentum operator along $z$. 
In the unmodulated phase, beyond a critical $\Omega>\Omega_c$, the solution 
with a vortex at the trap center becomes energetically favorable. In the 
presence of a density modulation, rotation does not only eventually result 
in the presence of vortices in the ground state, but may compromise as 
well the density pattern of the non-rotating ground state. In the D-SS phase, 
the ground state under rotation remains a droplet array~(although the number 
of droplets may change), presenting, beyond a critical rotation frequency, 
vortices in the inter-droplet interstitials \cite{Roccuzzo2020,Gallemi2020,Sindik2022}. 
Interestingly, the situation is different for other density patterns.

We focus our attention on the particular case of $N=2\times10^5$ atoms, which 
for the range of scattering lengths considered comprises all possible 
density patterns. Figure~\ref{fig:rotation_diagram} 
depicts the ground state phase diagram as a function of $\Omega$ and $a$. 
As mentioned above, the droplet array is robust under rotation. In contrast, 
both the stripe phase and the H-SS are not. Increasing 
the rotation frequency results in a transition of the ground-state from a 
non-rotating stripe or H-SS pattern into a labyrinthic phase with a vortex. 
There is hence no ground state with a vortex with a stripe or H-SS pattern. 

This has important consequences for the actual nucleation of vortices in 
experiments, which is typically realized by stirring the condensate by 
means of a slightly anisotropic rotating confinement on the $xy$ plane. 
Although the solution without vortex is not the ground state, it remains 
metastable for $\Omega>\Omega_c$ because moving a vortex to the center 
of the trap demands overcoming a potential barrier \cite{Sinha2001}. Such 
a barrier is eventually circumvented for a sufficiently large rotational 
frequency, typically much larger $\Omega_c$, at which the quadrupole surface mode 
is destabilized. This remains true also for a D-SS, although the 
critical rotation frequency for dynamical vortex nucleation may be significantly reduced 
compared to the unmodulated case~\cite{Gallemi2020}.

\begin{figure}[t!]
\centering
\includegraphics[width=\linewidth]{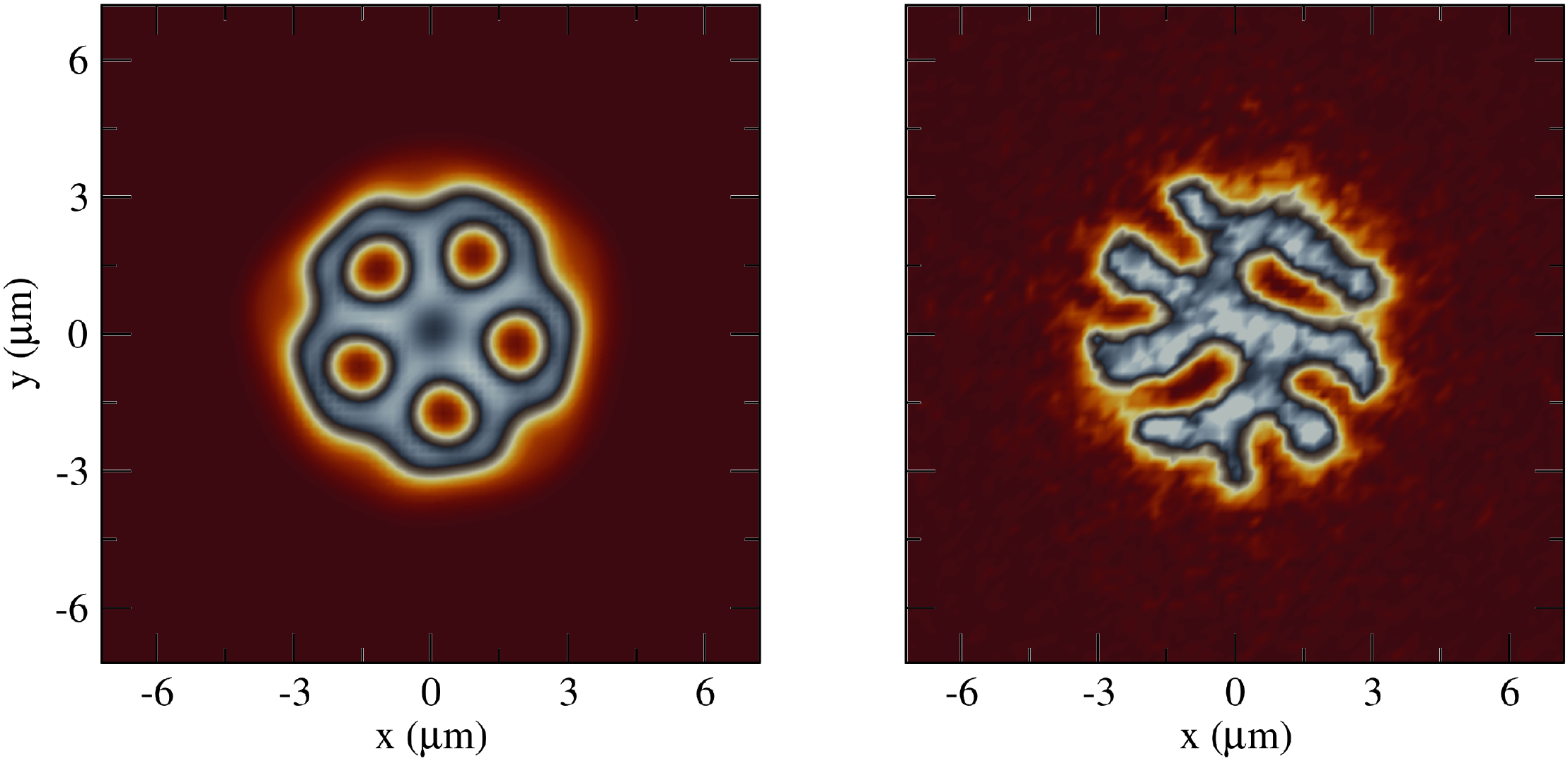}
\caption{(left panel) Initial condensate (without rotation) 
in the H-SS phase at $t=0$ for $a=82a_0$ (and the same parameters as 
Fig.~\ref{fig:moment_inertia}). When the trap is rotated with a trap anisotropy 
$\delta=0.025$~(see text) increasing adiabatically with a linear ramp 
of $0.03$ ms the rotation frequency up to 
$\Omega=0.9\omega_x$~(over the threshold frequency for the dynamical nucleation of vortices), 
the honeycomb pattern is destroyed into a labyrinthic arrangement, shown in the right panel 
at $t=43$ ms.}
\label{fig:vortex_breaking}
\end{figure}

In the D-SS phase, since the ground state under rotation remains a droplet array, 
the surface deformation is followed by the penetration of 
a vortex or vortices through the interstitials while keeping the droplet 
structure~\cite{Roccuzzo2020,Gallemi2020}. The situation is radically different 
in the stripe and H-SS patterns, due to the absence of a ground state with a 
vortex in those cases. When the condensate is stirred at a rotation frequency 
high-enough to destabilize the quadrupole surface mode, the stripe and the 
honeycomb patterns are destroyed while vortices nucleate. Therefore, as 
expected from the ground-state properties discussed above, a labyrinthic 
pattern is formed with vortices in its interstitials. 
This is illustrated in Fig.~\ref{fig:vortex_breaking}. Starting with the 
non-rotating H-SS ground state for $a=82a_0$~(left panel), we increase adiabatically 
$\Omega$. The eventual surface deformation for rotational frequencies 
above the quadrupole frequency leads to the creation of 
a vortex or vortices, but, prior to that, also to the destruction of 
the honeycomb pattern~(right panel).

As discussed below, the H-SS constitutes a much better supersolid 
than the droplet array. However, our results show that the lack of robustness of the density pattern under rotation is expected to prevent under typical experimental conditions 
the direct probe of superfluidity of that phase by means of the creation of 
quantized vortices. Fortunately, as shown in the following, the study 
of the moment of inertia constitutes a feasible alternative to directly probe the superfluidity of the H-SS.




\section{Moment of inertia}
\label{sec:momentinertiasec}

The moment of inertia may be theoretically evaluated from the 
response to an infinitesimally slow rotation, as 
$I_{\rm exact}=\lim_{\Omega \to 0} \frac{d\langle L_z\rangle}{d\Omega}$, 
where $\langle\hat{L}_z\rangle$ is the expected value of the angular 
momentum operator in the ground state calculated in the rotating frame.
The reduction of the moment of inertia with respect to its 
classical value in a cylindrically-symmetric quantum gas, 
$I_{\rm C}=m\int n(\mathbf{r})(x^2+y^2) d{\bf r}$, is a 
clear signature of a finite superfluid fraction, which together 
with the density modulation, becomes a direct probe for 
supersolidity~\cite{Leggett1970,Roccuzzo2020,Roccuzzo2022}. 

Figure~\ref{fig:moment_inertia}(a) shows, for $N=2\times10^5$, 
$I_{\rm exact}/I_{\rm C}$~(black dots) as a function of $a$ . 
The moment of inertia provides crucial information about the superfluid 
properties, although the lack of cylindrical symmetry may partially 
mask the true superfluid nature in some configurations. Due to the 
(quasi-)cylindrical symmetry of the unmodulated and stripe 
phases, the moment of inertia of these configurations turns out to 
be very close to zero. In both the labyrinth and the droplet 
patterns, the moment of inertia increases dramatically and approaches 
the classical rigid value, indicating the rigid response of the solid 
part. Most relevantly, the H-SS presents a very low moment 
of inertia, showing that, compared to the droplet array, it has typically 
a much larger superfluid fraction, while presenting a large density contrast. 


\section{Scissors-like perturbation}
\label{sec:scissors}

The reduction of the moment of inertia associated to superfluidity may be probed 
by monitoring the response of the system against a scissors-like perturbation. 
In the unmodulated phase, the frequency of the scissors mode may be directly 
linked to the superfluid fraction~\cite{GueryOdelin1999,Marago2000}. As shown 
by recent experiments on dipolar condensates, the situation is more subtle in 
the case of a D-SS~\cite{Tanzi2021,Norcia2022,Roccuzzo2022}, since determining 
the dominant frequency of the scissors mode does not generally allow for a clear 
proof of global superfluidity. However, a careful study of the overall spectrum 
of the system response against the scissors-like perturbation may provide an 
experimental estimation of the reduction of the moment of inertia under proper 
conditions~\cite{Roccuzzo2022}. A similar procedure may be employed for the study 
of superfluidity in the H-SS. 



\begin{figure}[t!]
\centering
\includegraphics[width=\linewidth]{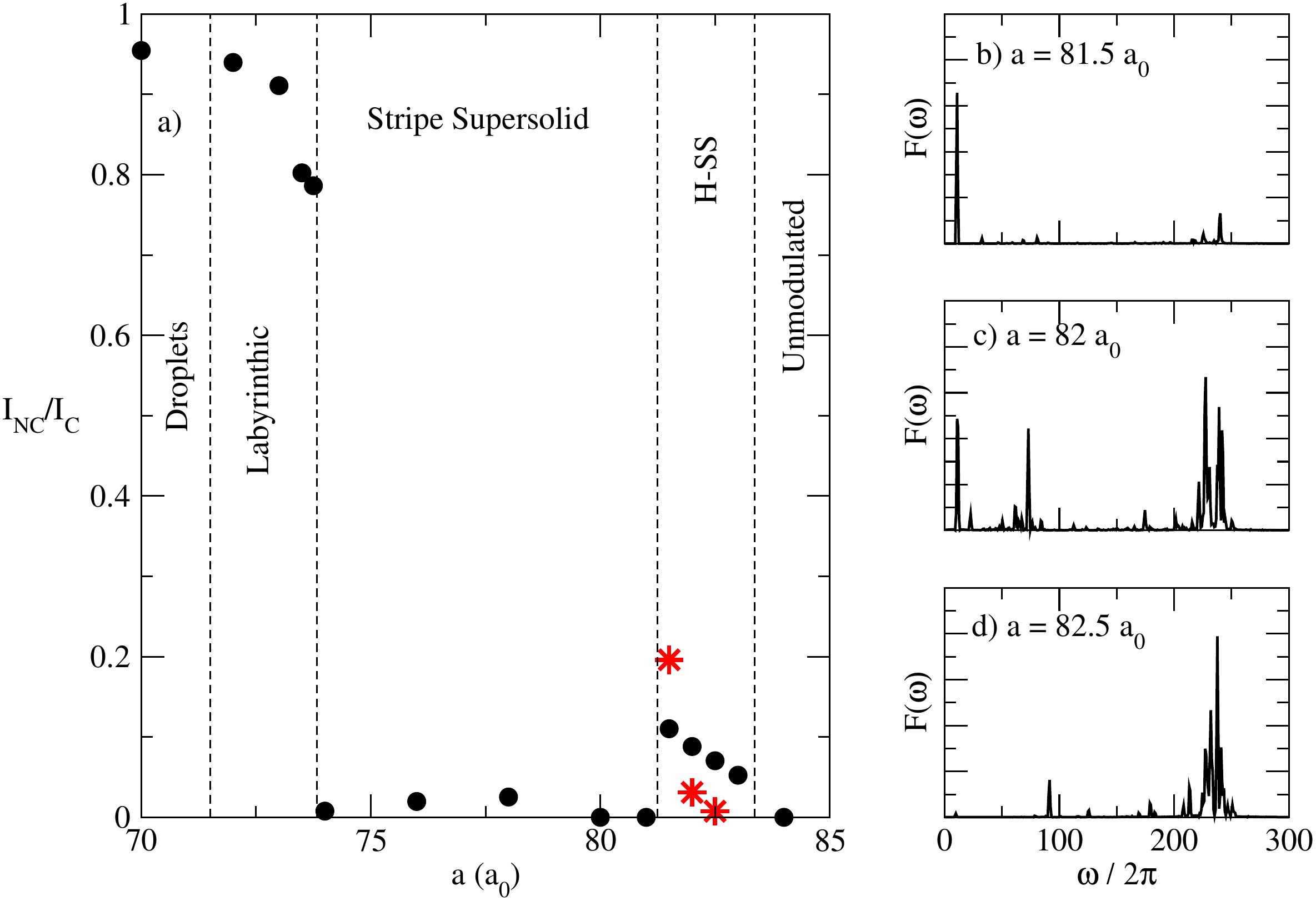}
\caption{(a) Moment of inertia $I_{\rm exact}/I_{\rm C}$~(black dots) for $N=2\times 10^5$ atoms, 
as a function of the scattering lengths~(all other parameters as in Fig.~\ref{fig:moment_inertia}). Panels 
(b)-(d) show the Fourier transform ${\cal F}(\omega)$ of 
the signal $\langle xy \rangle (t)$ of the scissors mode~(see text), measured after a sudden 
stop of a rotation of $1$ Hz for $a/a_0=81.5$, $82$, and $82.5$, respectively. 
The estimation of the moment of inertia using Eq.~\eqref{eq:Ip} for those cases is 
indicated with red stars in panel~(a). }
\label{fig:moment_inertia}
\end{figure}


We consider in the following a slightly deformed trap on the $xy$ plane, 
with $\omega_{x,y}=\omega(1\pm\delta)$, with $\delta=0.025$. Two possible 
experimental procedures may be employed for the study of the moment of 
inertia. The scissors-like perturbation may be induced either by starting 
with the ground state and tilting the trap on the $xy$ plane, or by rotating 
the condensate around $z$ and then suddenly stopping the rotation. In both cases, we 
may monitor the scissors-mode response $\langle xy \rangle (t)$ as a function 
of time and obtain the Fourier transform within a given time window $0<t<T$: 
${\cal F}(\omega)=\int_0^T dt \langle xy \rangle (t) e^{i\omega t}$. Using sum rules, it is 
possible to show that the moment of inertia may be estimated as~\cite{Roccuzzo2022}: 
\begin{equation}
I_{\rm p}=m(\omega_y^2-\omega_x^2)\langle x^2-y^2\rangle
\frac{\int d\omega \mathcal{F}(\omega)\omega^{p-2}}{\int d\omega \mathcal{F}(\omega)\omega^{p}}, 
\label{eq:Ip}
\end{equation}
with $p=0$~($p=1$) for the tilting~(rotation) protocol. 
Recent studies of the D-SS, have shown that 
both probing methods may in principle provide a reasonably good estimation 
of the moment of inertia. However, the tilting mechanism is particularly 
sensitive to low-frequency modes. As a result, that procedure demands 
prohibitively long integration times in order to provide reliable 
estimations of the moment of inertia. In contrast, the study of the scissors mode after 
suddenly stopping the rotation is much less sensitive to low frequencies, 
and hence provides a much better alternative.

Similar procedures may be employed to study experimentally the moment 
of inertia in the H-SS. We have evaluated for $N=2\times 10^5$ 
atoms the system response during $T=1$s~(the results are very consistent 
down to integration times as short as $200$ ms, well within realistic experimental lifetimes). 
The tilting procedure is very much affected by the very low-frequency modes 
characteristic of the polar quasi-symmetry of the H-SS pattern. As a result, that procedure 
provides a very bad estimation of $I_{\rm exact}/I_{\rm C}$; 
for $a=82a_0$, the estimation is $5.36$ versus the correct 
$0.08$ value. A much better result is obtained using the 
rotation technique~(see the red-star symbols, and panels (b) to (d) in 
Fig.~\ref{fig:moment_inertia}). Note that, although there is still a deviation from the actual value, the absolute difference with the exact result is small. Hence, the rotation method can be experimentally applied to reveal the much larger superfluid fraction of the H-SS compared to that of the D-SS.


\section{Conclusions}
\label{sec:conclusions}
Although research on dipolar supersolids has been limited up to now to the case of droplet 
supersolid arrays, experiments using larger number of atoms or tighter traps, or 
working with polar molecules, may open in the near future the possibility for studying other intriguing two-dimensional density patterns in quantum-stabilized dipolar condensates. Of those, honeycomb supersolids are particularly interesting, since, in contrast to droplet supersolids, their superfluid fraction remains very large even for a large density contrast. However, as discussed in this paper, the superfluidity of the honeycomb supersolid cannot be probed by the observation of quantized vortices due to the lack of robustness of the density pattern. Under rotation, vortex nucleation results in the destruction of the honeycomb pattern into a labyrinthic one. We have shown, however, that the large superfluid fraction of the honeycomb supersolid may be probed by a proper monitoring of the response after a scissors-like perturbation.


\section*{Acknowledgments}

We acknowledge support of the Deutsche Forschungsgemeinschaft 
(DFG, German Research Foundation) under Germany’s Excellence 
Strategy– EXC-2123 QuantumFrontiers–390837967, and FOR 2247. 

\bibliography{honeycomb_bib}

\end{document}